\pdfoutput=1
\documentclass[12pt]{article}
\usepackage{graphicx} %
\usepackage[margin=1in]{geometry}

\usepackage{hyperref} 
\usepackage{xcolor}
\usepackage{amsmath}
\usepackage{amsthm}
\usepackage{amsfonts}
\usepackage{booktabs}
\usepackage{multirow}
\usepackage[ruled,vlined]{algorithm2e}
\usepackage{natbib}
\usepackage{authblk}

\newcommand{\hL}{{\hat{L}}}

\newcommand{\bbE}{{\mathbb E}}

\newcommand{\bbR}{{\mathbb R}}

\newcommand{\cV}{{\mathcal{V}}}

\newcommand{\SL}{{\mathbb S}_L^2}
\newcommand{\SR}{{\mathbb S}_R^2}
\newcommand{\Sall}{{\mathbb S}^2}
\newcommand{\Data}{{\mathcal{D}}}
\newcommand{\DataC}{{\mathcal{D}_C}}

\newcommand{\diag}{\textrm{diag}}

\DeclareMathOperator*{\argmax}{arg\,max}

\newcommand{\Area}{{\textrm{Area}}}

\title{Bayesian Estimation of Continuous Brain Connectivity with Log-Gaussian Cox Processes 
}
\author[1]{Jaehoan Kim}
\author[2]{William Consagra}
\author[3]{Debdeep Pati}
\author[4]{Zhengwu Zhang}
\affil[1]{Department of Statistical Science, Duke University}
\affil[2]{Department of Statistics, University of South Carolina}
\affil[3]{Department of Statistics, University of Wisconsin-Madison}
\affil[4]{Department of Statistics and Operations Research, University of North Carolina at Chapel Hill}
\date{October 8, 2026}

\begin{document}

{\linespread{1}\selectfont\maketitle}

\section*{Abstract}

Continuous structural connectivity describes the connection between any two points of the cortical surface by an intensity function. Based on the endpoints of tractography streamlines, existing methods estimate this function mostly by kernel smoothing, which only gives a point estimate. We propose a Bayesian framework that estimates continuous connectivity with uncertainty quantification. The endpoints of tractography streamlines, mapped to two spheres, are modeled by a log-Gaussian Cox process on the product of the spheres with a Matérn Gaussian process prior. With the finite element representation of the prior, the Laplace approximation of the posterior enables the computation on a fine mesh with over $400{,}000$ latent variables per hemisphere pair. Numerical experiments show the accuracy of the point estimate of this approach and the coverage of its credible intervals. In the Adolescent Brain Cognitive Development Study, the proposed method predicts held-out streamlines better than kernel smoothing.

\section{Introduction}

The structural connectome is the network of white matter fibers that connect brain regions \citep{park2013structural, fornito2013graph}. Its organization is linked to cognitive abilities, and its alteration is associated with many neurological and psychiatric disorders \citep{craddock2013imaging, caznok2024between}. Diffusion MRI and tractography reconstruct these fibers non-invasively as a large set of streamlines \citep{zhang2022quantitative}, from which the structural connectivity is estimated.

The standard approach parcellates the cortex into regions of interest by a brain atlas \citep{desikan2006automated, glasser2016multi} and counts the streamlines between each pair of regions. The resulting connectivity matrix and the subsequent conclusions depend on the choice of the atlas \citep{zalesky2010whole, de2013parcellation, arslan2018human}, and averaging over large regions loses the spatial detail within each region \citep{moyer2017continuous}.

\begin{figure}[!htbp]
    \centering

    \begin{tabular}{c}
    (a) Atlas-based connectivity \\
    \includegraphics[width=0.8\linewidth]{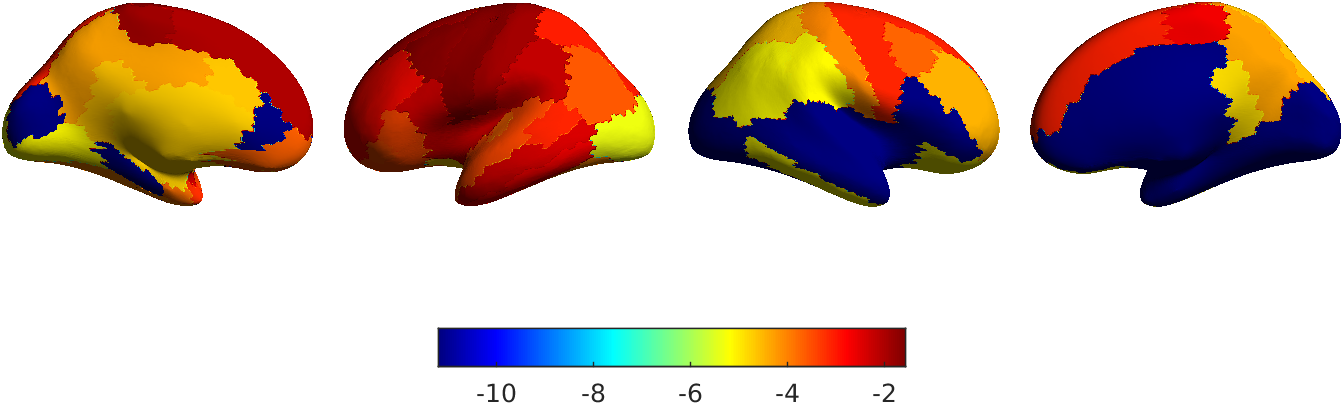} \\
    (b) Gaussian KDE, $\sigma = 0.046$ \\
    \includegraphics[width=0.8\linewidth]{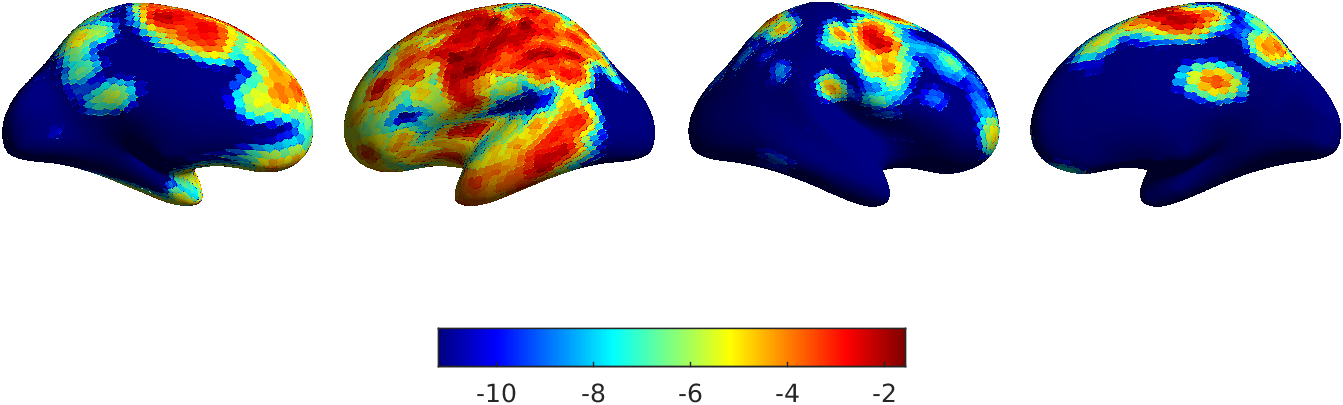} \\
    (c) Gaussian KDE, $\sigma = 0.184$ \\
    \includegraphics[width=0.8\linewidth]{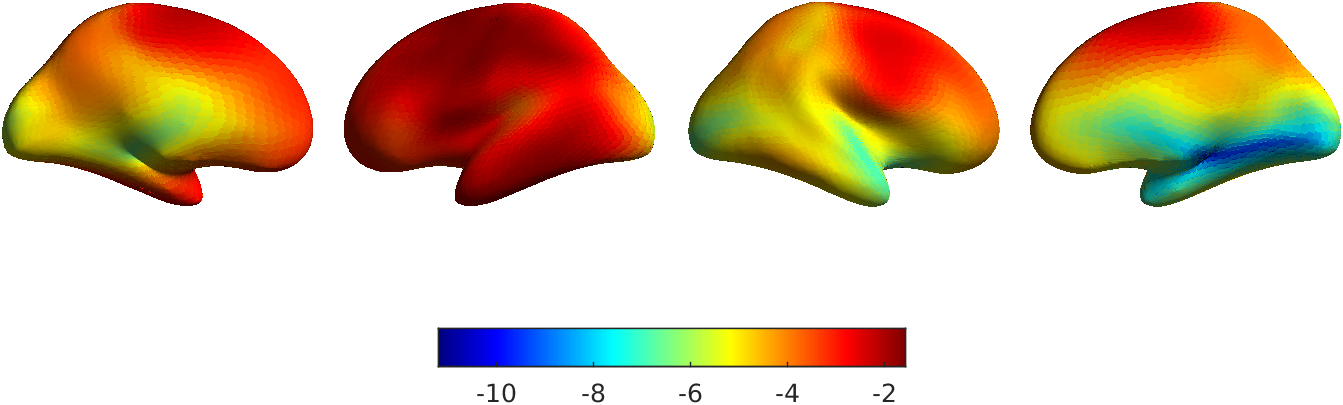}
    \end{tabular}
    \caption{Connectivity from the left precentral gyrus of one subject. (a) Streamline counts between the regions of the Desikan--Killiany atlas \citep{desikan2006automated}, divided by the region areas. (b, c) Gaussian KDE with the bandwidth selected by cross-validation and with four times this bandwidth. All panels show the normalized density on a common $\log_{10}$ scale, with the left and right hemispheres seen from both sides.}
    \label{fig:intro}
\end{figure}

Continuous structural connectivity avoids the parcellation \citep{moyer2017continuous, cole2021surface, seguin2022CSS}. Here, the streamline endpoints are modeled as a point process on the cortical surface, and the target is the intensity function $L(x, y)$ of the connections between any two points $x$ and $y$ of the cortex. This representation is more reproducible and predicts behavioral traits better than the region-based one \citep{cole2021surface, consagra2024continuous}. It also allows the joint alignment of the connectivity and the cortical surfaces across subjects \citep{cole2025alignment}. Existing methods estimate $L$ by kernel density estimation (KDE) \citep{moyer2017continuous, cole2021surface, seguin2022CSS}, with spherical kernels after the cortical surface is mapped to a sphere \citep{moyer2017continuous} or with geodesic distances on the cortical mesh \citep{seguin2022CSS}, or with the Riemannian diffusion kernel of the cortical surface \citep{wang2025riemannian}. \citet{consagra2026neuropmd} represent the density by a neural network on the product of two spheres and fit it by penalized maximum likelihood. These estimates depend on the choice of the kernel and the bandwidth or the penalty. Figure \ref{fig:intro} illustrates both issues for one subject in which the atlas removes the detail within each region, and the KDE changes substantially with the bandwidth. Moreover, these methods only give a point estimate, so one cannot assess the reliability of the estimate or test, for example, whether the connectivity differs between groups of subjects.

Gaussian process (GP) priors provide an alternative. The prior regularizes the estimate toward a smooth function, its smoothness is learned from the data through the hyperparameters, and the posterior quantifies the uncertainty of the estimate. However, the connectivity function is defined on a four-dimensional product of cortical surfaces \citep{consagra2024continuous}, and a high-resolution estimate needs a large number of discretization points. Standard GP computation scales cubically in this number \citep{rasmussen2006gaussian, liu2020gaussian}, which is infeasible at this scale.

For density estimation, the logistic GP prior places a GP on the log-density \citep{leonard1978density, lenk1991towards, tokdar2007towards}, and the log-Gaussian Cox process models a point pattern by a Poisson process whose log-intensity is a GP \citep{moller1998log}. Given the log-intensity and the number of points, the points of the Cox process are independent draws from the density proportional to the intensity, so the two models are closely related. Computation-wise, the latent function is discretized on a grid or a mesh, and the posterior is computed by Markov chain Monte Carlo \citep{tokdar2007towards, kim2026finite}, the Laplace approximation \citep{riihimaki2014laplace}, or the integrated nested Laplace approximation with the SPDE representation \citep{simpson2016going}. On a grid, the Kronecker structure of a separable covariance also gives fast Laplace approximations for non-Gaussian likelihoods \citep{saatci2012scalable, flaxman2015fast}. However, the number of latent variables on a grid grows exponentially with the dimension, which makes these approaches impractical beyond three or four dimensions \citep{riihimaki2014laplace}.

We propose a Bayesian framework for continuous structural connectivity based on a log-Gaussian Cox process on the product of two spheres. Through the link between Mat\'ern fields and stochastic partial differential equations (SPDEs) \citep{whittle1963stochastic, lindgren2011explicit}, the Mat\'ern GP prior on the log-intensity is represented by finite elements on a triangulation of the two spheres. Using the eigenbasis of the finite element matrices on each sphere, the Laplace approximation selects the hyperparameters and gives posterior draws using only products with the Fisher information, so no factorization of a large matrix is needed. This enables the computation on a mesh of the four-dimensional product with over $400{,}000$ latent variables per hemisphere pair. On a fine mesh, the proposed method gives more accurate point estimates than kernel density estimators in simulations and better held-out prediction in the Adolescent Brain Cognitive Development (ABCD) Study \citep{casey2018adolescent}, together with uncertainty quantification.

The rest of this paper is organized as follows. Section \ref{sec: model spec} presents the model, and Section \ref{sec: computation} describes the posterior computation. Sections \ref{sec: simulation} and \ref{sec: real data} present the simulation study and the application to the ABCD data, and Section \ref{sec: discussion} concludes the paper.

\section{Model}\label{sec: model spec}

Tractography applied to diffusion MRI reconstructs a large set of streamlines, and each streamline contributes an unordered pair of endpoints on the cortical surface \citep{st2018surface}. Following \citet{moyer2017continuous} and \citet{cole2021surface}, each hemisphere is inflated to a sphere, so that the endpoints lie on $\Omega = \SL \cup \SR$, with $\SL$ and $\SR$ denoting the left and the right hemisphere of the brain, respectively. We model the endpoint pairs $\Data = \{(x_i, y_i)\}_{i=1}^n$ as a realization of a Poisson point process on $\Omega \times \Omega$ with intensity $L = \exp(l)$, where $L(x, y) = L(y, x)$ as streamlines have no direction. The log-likelihood \citep[Chapter 3]{moller2004statistical} is written as
\begin{equation}\label{eqn: point process likelihood}
\log p(\Data \mid l) = \sum_{i=1}^n l(x_i, y_i) - \int_{\Omega \times \Omega} \exp\{l(x, y)\}\,dx\,dy - \log(n!).
\end{equation}
Since $\Omega$ is the disjoint union of two spheres, $\Omega \times \Omega$ splits into the blocks $\SL \times \SL$, $\SL \times \SR$, $\SR \times \SL$, and $\SR \times \SR$, and the symmetry of $L$ determines the block $\SR \times \SL$ from the block $\SL \times \SR$. We therefore model three log-intensities, $l_{LL}$ on $\SL \times \SL$, $l_{LR}$ on $\SL \times \SR$, and $l_{RR}$ on $\SR \times \SR$, where $l_{LL}$ and $l_{RR}$ are symmetric in their two arguments. Each inter-hemispheric streamline is assigned to $\SL \times \SR$ by ordering its endpoints so that $x_i \in \SL$ and $y_i \in \SR$.

As a Bayesian approach, we place on each $l_b$ a Mat\'ern Gaussian process prior with smoothness $\nu = 1$. Since the Mat\'ern field on $\bbR^d$ is the stationary solution of $(\kappa^2 - \Delta)^{(\nu + d/2)/2}\, l = \mathcal{W}$, we define the prior through the stochastic partial differential equation
\begin{equation}\label{eqn: product SPDE}
(\kappa^2 - \Delta)^{3/2}\, l = \mathcal{W}
\end{equation}
on the four-dimensional product of two spheres, where $\Delta$ is the Laplace--Beltrami operator of the product and $\mathcal{W}$ is Gaussian white noise \citep{whittle1963stochastic, lindgren2011explicit}. This construction yields a valid Mat\'ern-type covariance on the manifold without a geodesic distance \citep{borovitskiy2020matern}, and its finite element solution gives a finite-dimensional representation of $l_b$ \citep{lindgren2011explicit, kim2025adaptive}. %
Both spheres are triangulated by the same icosahedral mesh \citep{kenner2003geodesic} with vertices $\cV = \{v_1, \ldots, v_{n_v}\}$ and triangles $\{T_1, \ldots, T_{n_t}\}$, with areas $a_k = \Area(T_k)$. Let $\{\psi_i\}_{i=1}^{n_v}$ be the continuous piecewise linear basis on one sphere with knots at $\cV$, uniquely determined by $\psi_i(v_j) = 1\{i = j\}$. For $x$ in a triangle with vertices $v_{i_1}, v_{i_2}, v_{i_3}$, the only nonzero values $\psi_{i_1}(x), \psi_{i_2}(x), \psi_{i_3}(x)$ are the barycentric coordinates of $x$, the nonnegative weights with sum one such that $x = \sum_{m=1}^3 \psi_{i_m}(x)\, v_{i_m}$. For each block $b \in \{LL, LR, RR\}$ we write
\begin{equation}\label{eqn: basis expansion}
l_b(x, y) = \beta_b + \sum_{i=1}^{n_v} \sum_{j=1}^{n_v} (W_b)_{ij}\, \psi_i(x)\, \psi_j(y),
\end{equation}
where the intercept $\beta_b$ is the overall level of $l_b$ at each block and $W_b \in \bbR^{n_v \times n_v}$ models the variation across the domain.

Following \citet{lindgren2011explicit}, assigning the following prior on $W_b$ leads to the finite element GP prior on $l_b$ based on \eqref{eqn: product SPDE}:
\begin{equation}\label{eqn: FEM precision}
\mathrm{vec}(W_b) \sim \mathcal{N}(0, Q^{-1}),
\qquad
Q \propto K_\kappa C^{-1} K_\kappa C^{-1} K_\kappa,
\qquad
K_\kappa = \kappa^2 C + G,
\end{equation}
with $C = C_1 \otimes C_1$ and $G = C_1 \otimes G_1 + G_1 \otimes C_1$. Here $G_1$ is the stiffness matrix of the basis on one sphere, with entries $(G_1)_{ij} = \int_{\Sall} \langle \nabla \psi_i(x), \nabla \psi_j(x) \rangle\,dx$, and $C_1$ is the lumped mass matrix, the diagonal matrix with $(C_1)_{ii} = \int_{\Sall} \psi_i(x)\,dx = \sum_{k: v_i \in T_k} a_k / 3$, which replaces the mass matrix with entries $\int_{\Sall} \psi_i(x) \psi_j(x)\,dx$ by a diagonal matrix. See \citet{lindgren2011explicit} for the derivation of $Q$. The scale of $Q$ is set so that the area-averaged marginal variance of the field equals $\tau^2$ (Section \ref{sec: computation}). For the symmetric blocks the field is symmetrized so that $W_b = W_b^\top$, and the constant function, which lies in the null space of $G$, is represented by the intercept alone. The intercepts $\beta_b$ have independent weakly informative $\mathcal{N}(0, \sigma_\beta^2)$ priors. The hyperparameters $\theta = (\kappa, \tau)$ are shared by the three blocks, and $\kappa$ is defined on the unit sphere, so that the Mat\'ern range $\rho = \sqrt{8\nu}/\kappa$ is measured in radians and is comparable across meshes.

Next, we consider the likelihood function under this model. The integral in \eqref{eqn: point process likelihood} is evaluated on the mesh, and we therefore model the counts of endpoint pairs on the triangle pairs. For block $b$, let $(\DataC^b)_{kk'}$ be the number of endpoint pairs of the block in $T_k \times T_{k'}$, so that each streamline is counted once. We assume that the endpoint pairs of each block form independent Poisson processes. Then the counts in disjoint triangle pairs are independent Poisson variables with means
\begin{equation}\label{eqn: cell rate}
\mu^b_{kk'} = \int_{T_k} \int_{T_{k'}} \exp\{l_b(x, y)\}\,dx\,dy,
\end{equation}
and the log-likelihood of block $b$ is
\begin{equation}\label{eqn: count likelihood}
\log p(\DataC^b \mid l_b) = \sum_{k=1}^{n_t} \sum_{k'=1}^{n_t} \Big\{ (\DataC^b)_{kk'} \log \mu^b_{kk'} - \mu^b_{kk'} - \log\big((\DataC^b)_{kk'}!\big) \Big\}.
\end{equation}
The counts discard only the positions of the endpoints within their triangle pair. For the symmetric blocks, $\mu^b_{kk'} = \mu^b_{k'k}$, so \eqref{eqn: count likelihood} does not depend on how the within-hemisphere streamlines are oriented.

\section{Computation}\label{sec: computation}

Next, we describe the computation for the coefficients in our model in Section \ref{sec: model spec}. For notational simplicity, we consider parameters in one block and drop the block index. The three blocks are conditionally independent given $\theta = (\kappa, \tau)$, so they are fitted separately given $\theta$.

\subsection{Posterior distribution}\label{ssec: posterior}

The parameters of one block are the intercept $\beta$ and the coefficient matrix $W$ in \eqref{eqn: basis expansion}. We compute with the latent vector $u = (\mathrm{vec}(Z)^\top, \beta)^\top$ of dimension $n_v^2 + 1$, where $Z \in \bbR^{n_v \times n_v}$ is a whitened version of $W$ defined as follows. Let $C_1^{-1/2} G_1 C_1^{-1/2} = U \Lambda U^\top$ be the eigendecomposition of the normalized stiffness matrix, with eigenvalues $0 = \lambda_1 \le \lambda_2 \le \cdots \le \lambda_{n_v}$, and let $\Phi = C_1^{-1/2} U$. The columns of $\Phi$ are the discrete Laplace--Beltrami eigenfunctions on one sphere, normalized so that $\Phi^\top C_1 \Phi = I$. We write $W$ in this eigenbasis as
\begin{equation}\label{eqn: spectral representation}
W = \Phi\,\big(D_\theta \circ \tilde Z\big)\,\Phi^\top,
\end{equation}
where $\circ$ denotes the elementwise product, $\tilde Z = Z$ for the block $LR$ and $\tilde Z = (Z + Z^\top)/\sqrt{2}$ for the symmetric blocks $LL$ and $RR$, and
\begin{equation}\label{eqn: spectral scale}
(D_\theta)_{ab} = c_\theta\,(\kappa^2 + \lambda_a + \lambda_b)^{-3/2},
\qquad
c_\theta = \tau A_S \Big\{\sum_{a,b} (\kappa^2 + \lambda_a + \lambda_b)^{-3}\Big\}^{-1/2},
\end{equation}
for $(a, b) \neq (1, 1)$, with $(D_\theta)_{11} = 0$. Here $A_S = \mathrm{tr}(C_1)$ is the area of the sphere, and $c_\theta$ makes the area-averaged marginal variance of the Mat\'ern field equal to $\tau^2$ before its constant component is moved to the intercept. Since $K_\kappa = V \Sigma_\kappa V^\top$ with $V = C_1^{1/2}U \otimes C_1^{1/2}U$ and $\Sigma_\kappa = \kappa^2 I + \Lambda \otimes I + I \otimes \Lambda$, we have $Q^{-1} \propto (\Phi \otimes \Phi)\,\Sigma_\kappa^{-3}\,(\Phi \otimes \Phi)^\top$, and Appendix \ref{sec: spectral precision} shows that $W$ has the prior \eqref{eqn: FEM precision} when the entries of $Z$ are independent $\mathcal{N}(0, 1)$. The prior of $u$ is therefore $\mathcal{N}(0, P_0^{-1})$ with $P_0 = \diag(I_{n_v^2}, \sigma_\beta^{-2})$, which does not depend on $\theta$.

The likelihood \eqref{eqn: count likelihood} depends on $l$ only through the rates $\mu_{kk'}$, which have no closed form since $l$ is bilinear on $T_k \times T_{k'}$. Both spheres carry the same mesh, and on each triangle $T_k$ we use the three nodes $x_{k,p}$, $p = 1, 2, 3$, whose barycentric coordinates are $(2/3, 1/6, 1/6)$ and its permutations \citep{dunavant1985high}, and the product rule
\begin{equation}\label{eqn: nine point rule}
\mu_{kk'} \approx \frac{a_k a_{k'}}{9} \sum_{p=1}^3 \sum_{q=1}^3 \exp\{l(x_{k,p}, x_{k',q})\},
\end{equation}
where $x_{k,p}$ lies on the first sphere of the block and $x_{k',q}$ on the second. The rule is exact for integrands that are polynomials of degree at most two in each argument. Let $B \in \bbR^{3n_t \times n_v}$ be the matrix with entries $B_{(k,p), i} = \psi_i(x_{k,p})$, and let $M = B\Phi$. The rule \eqref{eqn: nine point rule} needs $l$ at the $9n_t^2$ node pairs of the block. By \eqref{eqn: basis expansion} and \eqref{eqn: spectral representation}, these values form the $3n_t \times 3n_t$ matrix
\begin{equation}\label{eqn: node log intensity}
E(u) = \beta\,\mathbf{1}\mathbf{1}^\top + M\,\big(D_\theta \circ \tilde Z\big)\,M^\top,
\end{equation}
whose rows index the $3n_t$ nodes on the first sphere and whose columns index those on the second. Thus, by \eqref{eqn: nine point rule}, the $n_t \times n_t$ matrix of rates can be written as
\begin{equation}\label{eqn: rate matrix}
\mu(u) = S\,\Psi(u)\,S^\top,
\qquad
\Psi(u) = (\omega\omega^\top) \circ \exp\{E(u)\},
\end{equation}
where $\omega_{(k,p)} = a_k/3$, the exponential is elementwise, and $S \in \{0, 1\}^{n_t \times 3n_t}$ has entries $S_{k, (k', p)} = 1\{k = k'\}$, so that $S = I_{n_t} \otimes \mathbf{1}_3^\top$ sums the three nodes of each triangle.
With the quadrature, the log-likelihood \eqref{eqn: count likelihood} becomes, up to a constant,
\begin{equation}\label{eqn: approximate likelihood}
\ell(u) = \sum_{k,k'} \big\{ (\DataC)_{kk'} \log \mu_{kk'}(u) - \mu_{kk'}(u) \big\}.
\end{equation}
Combined with the prior $u \sim \mathcal{N}(0, P_0^{-1})$, the posterior of $u$ given the data and $\theta$ is
\begin{equation}\label{eqn: log posterior}
p(u \mid \DataC, \theta) \propto \exp\{-h_\theta(u)\},
\qquad
h_\theta(u) = -\ell(u) + \tfrac{1}{2} u^\top P_0\, u.
\end{equation}

\subsection{Laplace approximation}\label{ssec: laplace}

We approximate the posterior \eqref{eqn: log posterior} by a Gaussian distribution at its mode, which is the Laplace approximation with the expected information.

Let $J_\mu = \partial\,\mathrm{vec}(\mu(u)) / \partial u^\top$ be the Jacobian of the rates. By \eqref{eqn: approximate likelihood}, the gradient of $h_\theta$ is
\begin{equation}\label{eqn: gradient}
\nabla h_\theta(u) = P_0\, u - J_\mu^\top \diag\{\mathrm{vec}(\mu(u))\}^{-1} \mathrm{vec}\big(\DataC - \mu(u)\big).
\end{equation}
Since the log-likelihood is not concave, we use the expected information in place of the Hessian,
\begin{equation}\label{eqn: hessian}
F(u) = \bbE\{\nabla^2 h_\theta(u)\} = P_0 + J_\mu^\top \diag\{\mathrm{vec}(\mu(u))\}^{-1} J_\mu,
\end{equation}
where the expectation is over $\DataC$ given $u$, which satisfies $F(u) \succeq P_0$. The mode $u^*$ is found by Fisher scoring \citep{mccullagh1989generalized}, in which each step solves $F(u)\,\delta = -\nabla h_\theta(u)$ by preconditioned conjugate gradients and is followed by a backtracking line search (Appendix \ref{sec: implementation}). Products with $F(u)$ are computed without forming $F(u)$, as described at the end of Section \ref{ssec: sampling}. The posterior is approximated by
\begin{equation}\label{eqn: gaussian approximation}
u \mid \DataC, \theta \approx \mathcal{N}\big(u^*,\, F(u^*)^{-1}\big).
\end{equation}

The procedure above approximates the posterior of $u$ for a given $\theta$. The hyperparameters, shared by the three blocks, are selected by maximizing the marginal likelihood,
\begin{equation}\label{eqn: empirical bayes}
\hat\theta = \argmax_{\theta} \sum_{b \in \{LL, LR, RR\}} \log p(\DataC^b \mid \theta),
\end{equation}
which is an empirical Bayes estimate under a flat prior on $(\log\kappa, \log\tau)$. The Laplace approximation gives each term as
\begin{equation}\label{eqn: laplace marginal likelihood}
\log p(\DataC \mid \theta)
\approx
\ell(u^*) - \tfrac{1}{2} (u^*)^\top P_0\, u^* - \tfrac{1}{2} \log \sigma_\beta^2 - \tfrac{1}{2} \log\det F(u^*) + \text{const}.
\end{equation}
The only expensive term is $\log\det F(u^*) = \mathrm{tr}\{\log F(u^*)\}$, which we estimate by stochastic Lanczos quadrature \citep{ubaru2017fast}. It averages $z_j^\top \log F(u^*)\, z_j$ over a few random vectors $z_j$ with independent $\pm 1$ entries, and each term is computed by the Lanczos method, which again needs only products with $F(u^*)$. The same $z_j$ are used for every $\theta$, so that \eqref{eqn: laplace marginal likelihood} is smooth in $\theta$. The maximization is carried out on $(\log\kappa, \log\tau)$ by a derivative-free search with warm starts of Fisher scoring (Appendix \ref{sec: implementation}). For the real data, a common $\hat\theta$ is selected on a grid for all subjects by summing the objective in \eqref{eqn: empirical bayes} over two scans. The procedure is a Laplace approximation with empirical Bayes hyperparameters in the spirit of integrated nested Laplace approximation \citep{rue2009approximate}, which is widely used for log-Gaussian Cox processes \citep{illian2012toolbox, simpson2016going}.

\subsection{Posterior sampling}\label{ssec: sampling}

Draws from \eqref{eqn: gaussian approximation} are obtained by perturbation-optimization \citep{papandreou2010gaussian}. Let $\epsilon_1 \sim \mathcal{N}(0, P_0)$, and let $\epsilon_2 \sim \mathcal{N}(0, I_{n_t^2})$. With $\mu$ evaluated at $u^*$, the vector $r = \epsilon_1 + J_\mu^\top \diag\{\mathrm{vec}(\mu)\}^{-1/2} \epsilon_2$ has covariance $F(u^*)$, so the solution $\delta$ of $F(u^*)\,\delta = r$ satisfies $\delta \sim \mathcal{N}(0, F(u^*)^{-1})$. Each draw $u^{(s)} = u^* + \delta^{(s)}$ requires one preconditioned conjugate gradient solve, and the draws are independent, so they can be computed in parallel. Pointwise posterior means and $95\%$ credible intervals are computed from $n_s$ draws. Algorithm \ref{alg: posterior computation} summarizes the whole procedure, from the selection of $\theta$ to the posterior draws.

\begin{algorithm}[!t]
\linespread{1}\selectfont
\caption{Posterior computation for continuous connectivity}
\label{alg: posterior computation}
\SetAlgoNoLine
\KwIn{Counts $\DataC^{LL}, \DataC^{LR}, \DataC^{RR}$ on the triangle pairs, mesh, number of draws $n_s$}
\KwOut{$\hat\theta$ and draws $l^{(1)}_b, \ldots, l^{(n_s)}_b$ of the log-intensities, $b \in \{LL, LR, RR\}$}
Compute $C_1$, $G_1$, the eigendecomposition $C_1^{-1/2} G_1 C_1^{-1/2} = U\Lambda U^\top$, $\Phi = C_1^{-1/2}U$, and $M = B\Phi$\;
\tcp{Selection of $\theta$ (Section \ref{ssec: laplace})}
\For{each candidate $\theta$ proposed by the search}{
  \For{$b \in \{LL, LR, RR\}$}{
    Compute the mode $u^*_b(\theta)$ of \eqref{eqn: log posterior} by Fisher scoring, warm started at the previous mode\;
    Estimate $\log\det F(u^*_b)$ by stochastic Lanczos quadrature and evaluate \eqref{eqn: laplace marginal likelihood}\;
  }
  Record $\sum_b \log p(\DataC^b \mid \theta)$\;
}
Set $\hat\theta$ by \eqref{eqn: empirical bayes}\;
\tcp{Posterior draws at $\hat\theta$ (Section \ref{ssec: sampling})}
\For{$b \in \{LL, LR, RR\}$}{
  Compute $u^*_b(\hat\theta)$\;
  \For{$s = 1, \ldots, n_s$}{
    Draw $\epsilon_1, \epsilon_2$, solve $F(u^*_b)\,\delta = r$ by conjugate gradients, and set $u^{(s)}_b = u^*_b + \delta = (\mathrm{vec}(Z^{(s)}_b)^\top, \beta^{(s)}_b)^\top$\;
    Set $W^{(s)}_b = \Phi\big(D_{\hat\theta} \circ \tilde Z^{(s)}_b\big)\Phi^\top$, which gives $l^{(s)}_b$ with intercept $\beta^{(s)}_b$ by \eqref{eqn: basis expansion}\;
  }
}
\end{algorithm}

Next, we consider the cost of Algorithm \ref{alg: posterior computation}. Its only matrix factorization is the eigendecomposition of the $n_v \times n_v$ matrix $C_1^{-1/2} G_1 C_1^{-1/2}$ on one sphere, which costs $O(n_v^3)$ once per mesh. All other computation consists of products with $F$, in the conjugate gradient iterations of Fisher scoring and of the draws and in the Lanczos steps, and of products with $J_\mu^\top$ in the gradient \eqref{eqn: gradient}. Neither $J_\mu$ nor $F$ is formed. Since $E(u)$ in \eqref{eqn: node log intensity} is linear in $u$, $J_\mu v = \mathrm{vec}[S\{\Psi(u) \circ E(v)\}S^\top]$, and $J_\mu^\top$ applies the transposes of the same maps in reverse order. A product with $F$ thus needs the transforms $\Phi X \Phi^\top$ and $\Phi^\top X \Phi$ of $n_v \times n_v$ matrices, at $O(n_v^3)$ operations, and products with $B$ and $S$. Since each row of $B$ and of $S$ has three nonzero entries, the latter are computed over the $n_t^2$ triangle pairs at $O(n_t^2)$ operations. On a triangulated sphere $n_t = 2n_v - 4$, so a product with $F$ costs $O(n_v^3) = O(N^{3/2})$ operations and $O(N)$ memory, where $N = n_v^2$ is the number of latent variables. A Cholesky factorization of the dense $N \times N$ matrix $F$ would instead cost $O(N^3)$ operations and $O(N^2)$ memory. Since $N = n_v^2$, this cost grows as $n_v^6$, so it becomes infeasible already for meshes with a few hundred vertices on each sphere.

\section{Simulation}\label{sec: simulation}

We now investigate the performance of the proposed method in a simulation study. We consider the following three true log-intensities $l_0$, generated independently for each block.
\begin{itemize}
\item Slowly varying GP. The coefficient matrix of $l_0$ is drawn exactly from the prior \eqref{eqn: FEM precision} through \eqref{eqn: spectral representation} on the ico3 mesh, with $\tau = 1$ and $\kappa = 2$, which gives the Mat\'ern range of $1.41$ radians.
\item Rapidly varying GP. The same with $\kappa = 6$, which gives the Mat\'ern range of $0.47$ radians.
\item von Mises--Fisher mixture.
\begin{equation*}
\exp\{l_0(x, y)\} \propto 0.05\,(4\pi)^{-2} + 0.95 \sum_{j=1}^{20} p_j\, g(x; m_j, \eta_j)\, g(y; m_j', \eta_j'),
\end{equation*}
where $g(\cdot\,; m, \eta)$ is the von Mises--Fisher density with center $m$ and concentration $\eta$. The centers are uniform on the sphere, $\eta_j, \eta_j' \in \{4, 16, 64\}$ with equal probabilities, and $(p_1, \ldots, p_{20})$ follows the flat Dirichlet distribution. For the blocks LL and RR, $\exp(l_0)$ is symmetrized in $(x, y)$.
\end{itemize}
The first two intensities fall into our model, whereas the third does not. The intensities are scaled so that the LL, LR, and RR blocks have $225{,}000$, $50{,}000$, and $225{,}000$ expected endpoint pairs, which matches the scale and the proportions of the real data. The endpoint pairs are drawn in two steps. First, the number of pairs in each cell pair $T_k \times T_{k'}$ of the ico3 mesh, which has $n_v = 642$ vertices and $n_t = 1{,}280$ triangles on each sphere, is Poisson with mean $\mu_{kk'}$ computed from $l_0$ by a quadrature finer than \eqref{eqn: nine point rule}, with six points in each triangle \citep{dunavant1985high}, and the pairs in the cell pair are drawn from the density proportional to $\exp(l_0)$ by rejection sampling.

We fit the proposed method on three nested meshes, ico1, ico2, and ico3, with $(n_v, n_t) = (42, 80)$, $(162, 320)$, and $(642, 1{,}280)$ and average edge lengths of $0.59$, $0.30$, and $0.15$ radians on each sphere. On ico1 and ico2, each endpoint is located in the coarse triangle that contains its radial projection. We set $\sigma_\beta = 10$ here and in Section \ref{sec: real data}. For each mesh, $\theta$ is selected by \eqref{eqn: empirical bayes}, and $50$ draws are taken from the Gaussian approximation \eqref{eqn: gaussian approximation}.

We compare with two kernel density estimators. For block $b$, the intensity estimate is
\begin{equation*}
\hL_b(x, y) = \sum_{i \in \mathcal{I}_b} K_\sigma(x, x_i)\, K_\sigma(y, y_i),
\end{equation*}
where $\mathcal{I}_b$ indexes the endpoint pairs of block $b$, and the estimate is symmetrized for the blocks LL and RR. The heat KDE of \citet{moyer2017continuous} uses the unit normalized spherical heat kernel $K_\sigma(x, x') = \sum_{h=0}^{\infty} (2h+1)(4\pi)^{-1} \exp\{-h(h+1)\sigma\}\, P_h(\langle x, x'\rangle)$ with the Legendre polynomials $P_h$, which we compute with the eigenpairs of the ico3 mesh as $\sum_{a=1}^{n_v} \exp(-\sigma\lambda_a)\, \phi_a(x)\, \phi_a(x')$, where $\phi_a = \sum_{i} \Phi_{ia} \psi_i$. The Gaussian KDE uses the kernel $\exp\{-d(x, x')^2 / (2\sigma^2)\}$ of the geodesic distance $d$ on the sphere, normalized to unit mass over the triangle centroids, which is nonnegative unlike the truncated heat kernel. For both, the endpoints are binned to the centroids of the ico3 triangles, and the bandwidth $\sigma$ is selected by five-fold likelihood cross-validation, in contrast to the integrated squared error criterion used in \citet{moyer2017continuous}. Namely, $\sigma$ maximizes $\sum_{m=1}^{5} \sum_{i \in \mathcal{F}_m} \log \hat f^{(-m)}(x_i, y_i)$ over a grid of $40$ log-spaced values, from $10^{-5}$ to $0.3$ for the heat kernel and from $0.005$ to $0.6$ for the Gaussian kernel, where $\mathcal{F}_m$ is the $m$th fold and $\hat f^{(-m)}$ is the density \eqref{eqn: intensity transformation to density} of the estimate from the other four folds. The same $\sigma$ is used for the three blocks, and the estimate is bounded below by $10^{-8}$ times its maximum before the logarithm is taken.
The three methods are compared through the normalized density. Let $\Omega_{LL} = \SL \times \SL$, $\Omega_{LR} = \SL \times \SR$, and $\Omega_{RR} = \SR \times \SR$. For the intensities $L_b$ of the three blocks, the density is
\begin{equation}\label{eqn: intensity transformation to density}
f_b(x, y) = \frac{L_b(x, y)}{\sum_{b'} \int_{\Omega_{b'}} L_{b'}},
\qquad (x, y) \in \Omega_b,
\end{equation}
where $L_b = \exp(l_b)$ for the proposed method. Let $f_0$ be the true density. For the proposed method, the estimate $\hat f$ is the posterior mean of the density over the draws, and the $95\%$ pointwise credible intervals are formed from the same draws. We measure the accuracy of $\hat f$ by
\begin{equation*}
\|\hat f - f_0\|_2 = \Big\{\sum_b \int_{\Omega_b} (\hat f_b - f_{0,b})^2\Big\}^{1/2},
\qquad
d_H(\hat f, f_0) = \Big\{\frac{1}{2} \sum_b \int_{\Omega_b} \big(\hat f_b^{1/2} - f_{0,b}^{1/2}\big)^2\Big\}^{1/2},
\end{equation*}
both computed from the averages of the densities over the ico3 cell pairs. For the credible intervals, we report the proportion of ico3 cell pairs whose true average density lies in the interval.

\begin{table}[ht]
    \centering
    \linespread{1}\selectfont
    \caption{Performance comparison across resolutions. Cov. denotes the empirical coverage (\%) of the $95\%$ pointwise posterior credible intervals of the true density. The columns correspond to the three true intensities. Results are from a single simulated data set.}
    \label{tab:performance}
    \vspace{4pt}
    \small\setlength{\tabcolsep}{4pt}
    \begin{tabular}{lccccccccc}
        \toprule
        & \multicolumn{3}{c}{$\kappa = 2$} & \multicolumn{3}{c}{$\kappa = 6$} & \multicolumn{3}{c}{vMF} \\
        \cmidrule(lr){2-4} \cmidrule(lr){5-7} \cmidrule(lr){8-10}
        Method & $L_2$ & Hellinger & Cov. & $L_2$ & Hellinger & Cov. & $L_2$ & Hellinger & Cov. \\
        \midrule
        heat KDE & 0.0153 & 0.0829 & -- & 0.0212 & 0.1245 & -- & 0.1789 & 0.1381 & -- \\
        Gaussian KDE & 0.0160 & 0.0873 & -- & 0.0209 & 0.1266 & -- & 0.1687 & 0.1382 & -- \\
        GP (ico1) & 0.0259 & 0.1355 & 58.4 & 0.0380 & 0.2179 & 39.5 & 2.8938 & 0.4703 & 35.3 \\
        GP (ico2) & 0.0139 & 0.0771 & 90.1 & 0.0233 & 0.1363 & 83.2 & 0.0853 & 0.0858 & 95.0 \\
        GP (ico3) & \textbf{0.0115} & \textbf{0.0678} & 90.5 & \textbf{0.0173} & \textbf{0.1082} & 90.3 & \textbf{0.0246} & \textbf{0.0568} & 96.3 \\
        \bottomrule
    \end{tabular}
\end{table}

Table \ref{tab:performance} shows the results of the GP estimator and the two kernel estimators. For both slowly varying and rapidly varying intensity function, the error of the GP estimator decreases as the mesh is refined, and GP under ico3 attains the smallest $L_2$ and Hellinger errors, below those of both kernel estimators, which perform similarly. The ico1 grid is too coarse to capture the behavior of the intensity function, which leads to a large bias and low coverage. The coarse resolution is more harmful for the rapidly varying function, for which GP under ico2 is less accurate than the kernel estimators. Under ico3, the $95\%$ credible intervals cover about $90\%$ of the true density values in both cases.

Under the von Mises--Fisher mixture, where the model does not hold, the error of the GP estimator again decreases as the mesh is refined. GP under ico2 and ico3 is much more accurate than both kernel estimators, and its credible intervals have coverage close to the nominal level. The kernel estimators use one bandwidth for the whole function, which cannot fit the broad and the sharp bumps at the same time. The GP estimator smooths on the log scale, and its posterior follows the data where the counts are large and shrinks to a smooth surface where they are small, so the strong and the weak connections are both recovered. GP under ico1 again fails because of its coarse resolution. Most of its large $L_2$ error comes from fewer than $0.1\%$ of the cell pairs, where the exponential of the linear interpolation over a coarse triangle concentrates the mass near a vertex.

\section{Real data application}\label{sec: real data}

We apply the proposed method to the structural connectivity of children in the Adolescent Brain Cognitive Development (ABCD) study \citep{casey2018adolescent}. The data contain $98$ subjects, each scanned at two time points, the baseline and the 2-year follow-up. The subjects were $9$ to $11$ years old at baseline, and $49$ of them were female. Streamlines were reconstructed by tractography, and their endpoints were mapped to the inflated spheres by the surface-based connectivity integration pipeline \citep{cole2021surface}. A scan contains a median of $493{,}340$ streamlines, and about $44\%$, $11\%$, and $44\%$ belong to the blocks LL, LR, and RR, respectively.

For each scan, we split the streamlines at random into two halves, estimate the connectivity from one half, and evaluate the estimate on the other half and on the scan at the other time point (Appendix \ref{sec: real data measures}). All fits use the ico3 mesh. A common $\theta$ is selected by maximizing the sum of \eqref{eqn: laplace marginal likelihood} over two scans on the grid $\kappa \in \{8, 14, 24, 40\}$ and $\tau \in \{3, 5, 8\}$, and the selected $\hat\theta = (24, 5)$ is used for all scans. For each scan, $20$ posterior draws are taken as in Section \ref{sec: simulation}, and the posterior mode is used as the point estimate. With $\hat\theta$ fixed, the fit of the three blocks of one scan, including the draws, takes about four to five hours on $8$ CPU cores. We compare with the heat and Gaussian kernel estimators of Section \ref{sec: simulation}, whose bandwidths are selected by five-fold likelihood cross-validation within the fitted half.

Table \ref{tab: real data} shows the predictive accuracy of the three methods. The proposed method predicts the held-out half better than both kernel estimators in every scan. Since a difference in the log-score estimates a difference in the Kullback--Leibler divergence from the true distribution of the ico3 region pairs (Appendix \ref{sec: real data measures}), the proposed estimate is closer to the true connectivity of each scan at this resolution. Across time points, the estimate from one scan of a subject also predicts the other scan of the subject better in almost all comparisons ($195$ and $189$ out of $196$ with the Gaussian and heat kernel estimators). This score is lowered both by noise in an estimate, which does not recur in the other scan, and by oversmoothing, which places mass where the other scan has few streamlines. Therefore, the proposed method captures more of the connectivity that is reproduced across scans.

\begin{table}[ht]
    \centering
    \linespread{1}\selectfont
    \caption{Real data. Average log-score per streamline (higher is better) on the held-out half of each scan and on the scan at the other time point. The two scores are not comparable, since the cross-session score is computed on the coarser ico2 region pairs.}
    \label{tab: real data}
    \vspace{4pt}
    \begin{tabular}{lcc}
        \toprule
        Method & Held-out & Cross-session \\
        \midrule
        GP (proposed) & $\mathbf{-11.560}$ & $\mathbf{-9.800}$ \\
        Gaussian KDE & $-11.668$ & $-9.877$ \\
        heat KDE & $-12.206$ & $-9.898$ \\
        \bottomrule
    \end{tabular}
\end{table}

The resolution of the mesh matters as in the simulation. When the model is fitted on ico2 to $20$ scans of $10$ subjects, the held-out log-score falls far below both the ico3 fit and the kernel estimators.

\begin{figure}[!htbp]
    \centering
    \begin{tabular}{c}
    (a) Gaussian KDE \\
    \includegraphics[width=0.8\linewidth]{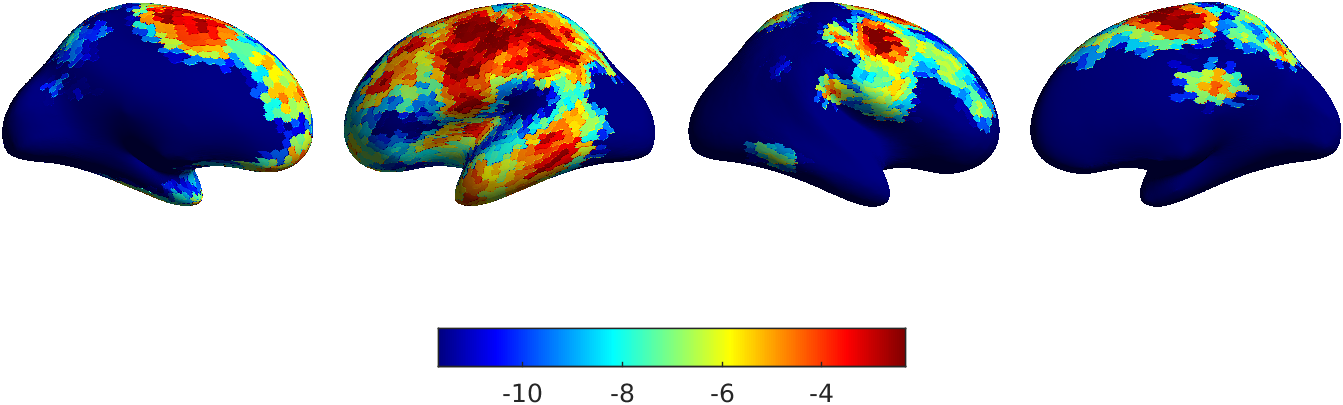} \\
    (b) GP posterior mode \\
    \includegraphics[width=0.8\linewidth]{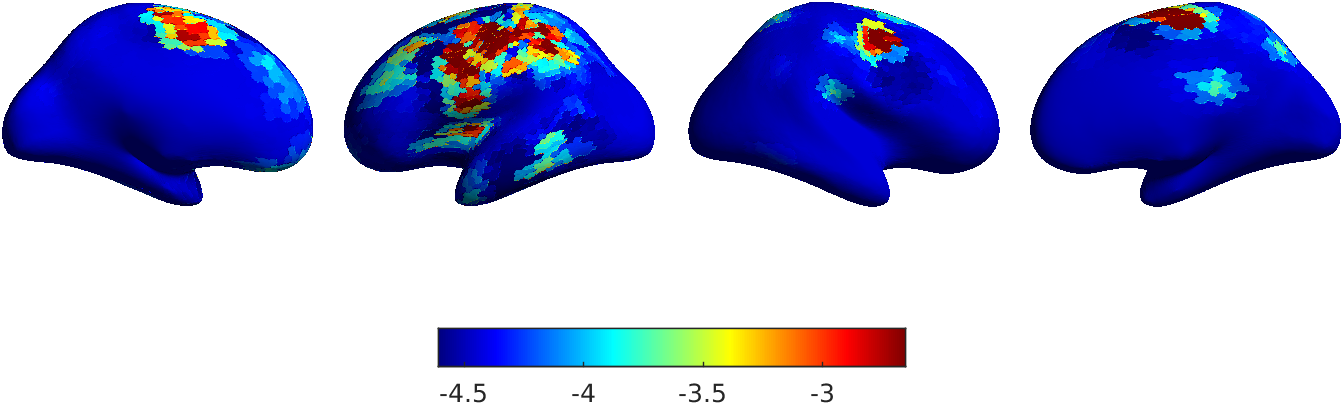} \\
    (c) GP posterior probability that $f$ exceeds its uniform level \\
    \includegraphics[width=0.8\linewidth]{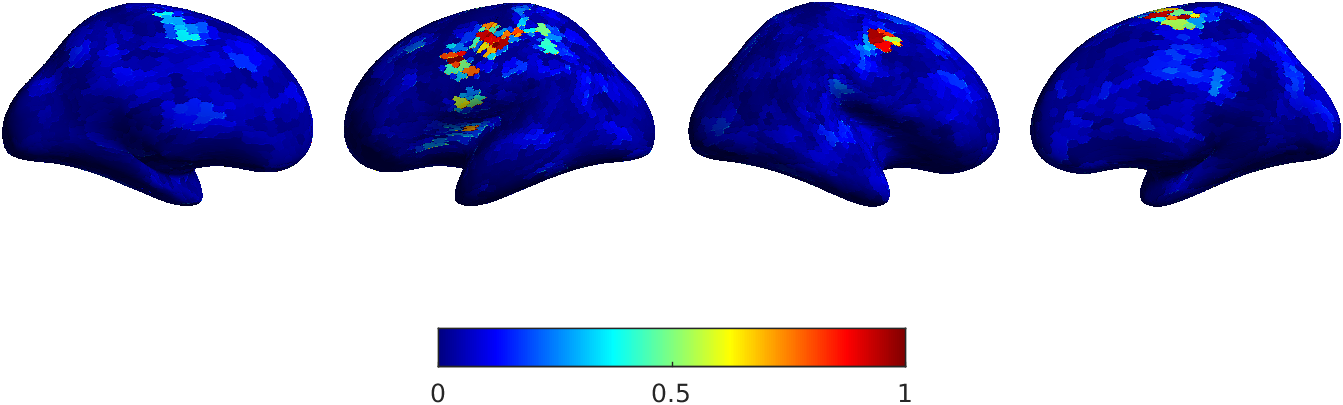}
    \end{tabular}
    \caption{Connectivity from a seed at the center of the left precentral gyrus for one subject (baseline) on the inflated fsaverage surface. In each panel, the first two views show the left hemisphere and the last two show the right hemisphere, each seen from both sides. Panels (a) and (b) show $\log_{10} f$ with separate color scales, and panel (c) shows the posterior probability that $f$ exceeds its value $1/(3A_S^2)$ under uniform connectivity.}
    \label{fig:brain_new}
\end{figure}

Figure \ref{fig:brain_new} shows the connectivity of one subject from a seed in the left precentral gyrus. The posterior mode and the kernel estimate place the strong connections at the same locations, and the posterior mode is smoother. Panel (c) shows the posterior probability that $f$ exceeds its value under uniform connectivity. In the regions of the Desikan--Killiany atlas \citep{desikan2006automated}, this probability is close to one (red) mostly in the precentral, postcentral and paracentral gyri of both hemispheres, which form the sensorimotor cortex. Therefore, the connections of the seed to the sensorimotor cortex of both hemispheres are strong beyond the uncertainty of the estimate.

Next, we check whether this uncertainty is consistent with the variation between the two scans of a subject. Let $m^{(t)}_{j,r}$ and $s^{(t)}_{j,r}$ be the posterior mean and standard deviation of the log mass of the ico2 region pair $r$ for subject $j$ at time $t$. If the connectivity does not change between the time points, $z_{j,r} = (m^{(1)}_{j,r} - m^{(2)}_{j,r}) / \{(s^{(1)}_{j,r})^2 + (s^{(2)}_{j,r})^2\}^{1/2}$ is approximately standard normal. Over all subjects and region pairs, $97.6\%$ of the values satisfy $|z_{j,r}| < 1.96$.

\section{Conclusion}\label{sec: discussion}

This paper introduces a Bayesian framework that estimates continuous structural connectivity with uncertainty quantification. Existing methods estimate the connectivity function mostly by kernel smoothing, which gives only a point estimate and depends on the choice of the kernel and the bandwidth. To address this, we model the streamline endpoints by a log-Gaussian Cox process on the product of two spheres with a Mat\'ern Gaussian process prior defined through an SPDE. Computation-wise, the Laplace approximation selects the hyperparameters and gives posterior draws on a mesh with over $400{,}000$ latent variables per hemisphere pair, using only products with the Fisher information. Numerical experiments show that the proposed estimate on a fine mesh is more accurate than kernel density estimators, and in the ABCD data it predicts held-out streamlines better in every scan, with uncertainty consistent with the variation between the two scans of each subject.

Several limitations remain. The posterior describes the uncertainty given the tractography output under the Poisson model, so it does not include the variation from tractography, such as the seeding and the dependence between streamlines. The Mat\'ern range selected for the real data is comparable to the edge length of the ico3 mesh, so a finer mesh may further improve the estimate at a higher computational cost. The Kronecker structure needs the same mesh on the two spheres, so the method does not directly apply to two different cortical surface meshes. Lastly, integrating over $\theta$ instead of fixing it at $\hat\theta$ would add the uncertainty of the hyperparameters to the posterior. The posterior draws of each subject can be carried to group-level analyses, such as testing the difference in connectivity between groups of subjects.

\section*{Acknowledgments}

\begin{sloppypar}
Data used in the preparation of this article were obtained from the Adolescent Brain Cognitive Development (ABCD) Study (\url{https://abcdstudy.org}), held in the NIMH Data Archive (NDA). This is a multisite, longitudinal study designed to recruit more than 10,000 children age 9--10 and follow them over 10 years into early adulthood. The ABCD Study is supported by the National Institutes of Health and additional federal partners under award numbers U01DA041048, U01DA050989, U01DA051016, U01DA041022, U01DA051018, U01DA051037, U01DA050987, U01DA041174, U01DA041106, U01DA041117, U01DA041028, U01DA041134, U01DA050988, U01DA051039, U01DA041156, U01DA041025, U01DA041120, U01DA051038, U01DA041148, U01DA041093, U01DA041089, U24DA041123, U24DA041147. A full list of supporters is available at \url{https://abcdstudy.org/federal-partners.html}. A listing of participating sites and a complete listing of the study investigators can be found at \url{https://abcdstudy.org/consortium_members/}. ABCD consortium investigators designed and implemented the study and/or provided data but did not necessarily participate in analysis or writing of this report. This manuscript reflects the views of the authors and may not reflect the opinions or views of the NIH or ABCD consortium investigators. The ABCD data repository grows and changes over time. The ABCD data used in this report came from NIMH Data Archive Digital Object Identifier (\href{https://dx.doi.org/10.15154/1520158}{10.15154/1520158}).
\end{sloppypar}

\bibliographystyle{plainnat}
\bibliography{ref}

\begin{thebibliography}{43}
\providecommand{\natexlab}[1]{#1}
\providecommand{\url}[1]{\texttt{#1}}
\expandafter\ifx\csname urlstyle\endcsname\relax
  \providecommand{\doi}[1]{doi: #1}\else
  \providecommand{\doi}{doi: \begingroup \urlstyle{rm}\Url}\fi

\bibitem[Arslan et~al.(2018)Arslan, Ktena, Makropoulos, Robinson, Rueckert, and
  Parisot]{arslan2018human}
Salim Arslan, Sofia~Ira Ktena, Antonios Makropoulos, Emma~C Robinson, Daniel
  Rueckert, and Sarah Parisot.
\newblock Human brain mapping: A systematic comparison of parcellation methods
  for the human cerebral cortex.
\newblock \emph{NeuroImage}, 170:\penalty0 5--30, 2018.

\bibitem[Borovitskiy et~al.(2020)Borovitskiy, Terenin, Mostowsky, and
  Deisenroth]{borovitskiy2020matern}
Viacheslav Borovitskiy, Alexander Terenin, Peter Mostowsky, and Marc~Peter
  Deisenroth.
\newblock Mat{\'e}rn {G}aussian processes on {R}iemannian manifolds.
\newblock \emph{Advances in Neural Information Processing Systems},
  33:\penalty0 12426--12437, 2020.

\bibitem[Casey et~al.(2018)Casey, Cannonier, Conley, Cohen, Barch, Heitzeg,
  Soules, Teslovich, Dellarco, Garavan, et~al.]{casey2018adolescent}
Betty~Jo Casey, Tariq Cannonier, May~I Conley, Alexandra~O Cohen, Deanna~M
  Barch, Mary~M Heitzeg, Mary~E Soules, Theresa Teslovich, Danielle~V Dellarco,
  Hugh Garavan, et~al.
\newblock The {A}dolescent {B}rain {C}ognitive {D}evelopment ({ABCD}) {S}tudy:
  imaging acquisition across 21 sites.
\newblock \emph{Developmental Cognitive Neuroscience}, 32:\penalty0 43--54,
  2018.

\bibitem[Caznok~Silveira et~al.(2024)Caznok~Silveira, Antunes, Athi{\'e},
  da~Silva, Ribeiro~dos Santos, Canateli, Fontoura, Pinto, Pimentel-Silva,
  Avansini, et~al.]{caznok2024between}
Ana~Clara Caznok~Silveira, Andre Saraiva Le{\~a}o~Marcelo Antunes, Maria
  Carolina~Pedro Athi{\'e}, B{\'a}rbara~Filomena da~Silva, Jo{\~a}o~Victor
  Ribeiro~dos Santos, Camila Canateli, Marina~Alves Fontoura, Allan Pinto,
  Luciana~Ramalho Pimentel-Silva, Simoni~Helena Avansini, et~al.
\newblock Between neurons and networks: investigating mesoscale brain
  connectivity in neurological and psychiatric disorders.
\newblock \emph{Frontiers in Neuroscience}, 18:\penalty0 1340345, 2024.

\bibitem[Cole et~al.(2021)Cole, Murray, St-Onge, Risk, Zhong, Schifitto,
  Descoteaux, and Zhang]{cole2021surface}
Martin Cole, Kyle Murray, Etienne St-Onge, Benjamin Risk, Jianhui Zhong,
  Giovanni Schifitto, Maxime Descoteaux, and Zhengwu Zhang.
\newblock Surface-based connectivity integration: An atlas-free approach to
  jointly study functional and structural connectivity.
\newblock \emph{Human Brain Mapping}, 42\penalty0 (11):\penalty0 3481--3499,
  2021.

\bibitem[Cole et~al.(2025)Cole, Xiang, Consagra, Srivastava, Qiu, and
  Zhang]{cole2025alignment}
Martin Cole, Yang Xiang, William Consagra, Anuj Srivastava, Xing Qiu, and
  Zhengwu Zhang.
\newblock Alignment of continuous brain connectivity.
\newblock \emph{arXiv preprint arXiv:2503.15830}, 2025.

\bibitem[Consagra et~al.(2024)Consagra, Cole, Qiu, and
  Zhang]{consagra2024continuous}
William Consagra, Martin Cole, Xing Qiu, and Zhengwu Zhang.
\newblock Continuous and atlas-free analysis of brain structural connectivity.
\newblock \emph{The Annals of Applied Statistics}, 18\penalty0 (3):\penalty0
  1815--1839, 2024.

\bibitem[Consagra et~al.(2026)Consagra, Gu, and Zhang]{consagra2026neuropmd}
William Consagra, Zhiling Gu, and Zhengwu Zhang.
\newblock {NeuroPMD}: Neural fields for density estimation on product
  manifolds.
\newblock \emph{Journal of Computational and Graphical Statistics}, 35\penalty0
  (3):\penalty0 1277--1290, 2026.

\bibitem[Craddock et~al.(2013)Craddock, Jbabdi, Yan, Vogelstein, Castellanos,
  Di~Martino, Kelly, Heberlein, Colcombe, and Milham]{craddock2013imaging}
R~Cameron Craddock, Saad Jbabdi, Chao-Gan Yan, Joshua~T Vogelstein, F~Xavier
  Castellanos, Adriana Di~Martino, Clare Kelly, Keith Heberlein, Stan Colcombe,
  and Michael~P Milham.
\newblock Imaging human connectomes at the macroscale.
\newblock \emph{Nature Methods}, 10\penalty0 (6):\penalty0 524--539, 2013.

\bibitem[de~Reus and Van~den Heuvel(2013)]{de2013parcellation}
Marcel~A de~Reus and Martijn~P Van~den Heuvel.
\newblock The parcellation-based connectome: limitations and extensions.
\newblock \emph{NeuroImage}, 80:\penalty0 397--404, 2013.

\bibitem[Desikan et~al.(2006)Desikan, S{\'e}gonne, Fischl, Quinn, Dickerson,
  Blacker, Buckner, Dale, Maguire, Hyman, et~al.]{desikan2006automated}
Rahul~S Desikan, Florent S{\'e}gonne, Bruce Fischl, Brian~T Quinn, Bradford~C
  Dickerson, Deborah Blacker, Randy~L Buckner, Anders~M Dale, R~Paul Maguire,
  Bradley~T Hyman, et~al.
\newblock An automated labeling system for subdividing the human cerebral
  cortex on {MRI} scans into gyral based regions of interest.
\newblock \emph{NeuroImage}, 31\penalty0 (3):\penalty0 968--980, 2006.

\bibitem[Dunavant(1985)]{dunavant1985high}
D.~A. Dunavant.
\newblock High degree efficient symmetrical {G}aussian quadrature rules for the
  triangle.
\newblock \emph{International Journal for Numerical Methods in Engineering},
  21\penalty0 (6):\penalty0 1129--1148, 1985.

\bibitem[Flaxman et~al.(2015)Flaxman, Wilson, Neill, Nickisch, and
  Smola]{flaxman2015fast}
Seth Flaxman, Andrew~Gordon Wilson, Daniel~B. Neill, Hannes Nickisch, and
  Alexander~J. Smola.
\newblock Fast {K}ronecker inference in {G}aussian processes with
  non-{G}aussian likelihoods.
\newblock In \emph{Proceedings of the 32nd International Conference on Machine
  Learning}, volume~37 of \emph{Proceedings of Machine Learning Research},
  pages 607--616, 2015.

\bibitem[Fornito et~al.(2013)Fornito, Zalesky, and
  Breakspear]{fornito2013graph}
Alex Fornito, Andrew Zalesky, and Michael Breakspear.
\newblock Graph analysis of the human connectome: promise, progress, and
  pitfalls.
\newblock \emph{NeuroImage}, 80:\penalty0 426--444, 2013.

\bibitem[Glasser et~al.(2016)Glasser, Coalson, Robinson, Hacker, Harwell,
  Yacoub, Ugurbil, Andersson, Beckmann, Jenkinson, et~al.]{glasser2016multi}
Matthew~F Glasser, Timothy~S Coalson, Emma~C Robinson, Carl~D Hacker, John
  Harwell, Essa Yacoub, Kamil Ugurbil, Jesper Andersson, Christian~F Beckmann,
  Mark Jenkinson, et~al.
\newblock A multi-modal parcellation of human cerebral cortex.
\newblock \emph{Nature}, 536\penalty0 (7615):\penalty0 171--178, 2016.

\bibitem[Illian et~al.(2012)Illian, S{\o}rbye, and Rue]{illian2012toolbox}
Janine~B Illian, Sigrunn~H S{\o}rbye, and H{\aa}vard Rue.
\newblock A toolbox for fitting complex spatial point process models using
  integrated nested {L}aplace approximation ({INLA}).
\newblock \emph{The Annals of Applied Statistics}, 6\penalty0 (4):\penalty0
  1499--1530, 2012.

\bibitem[Kenner(2003)]{kenner2003geodesic}
Hugh Kenner.
\newblock \emph{Geodesic math and how to use it}.
\newblock Univ of California Press, 2003.

\bibitem[Kim et~al.(2025)Kim, Bhattacharya, and Pati]{kim2025adaptive}
Jaehoan Kim, Anirban Bhattacharya, and Debdeep Pati.
\newblock Adaptive resolution for finite-rank {G}aussian processes.
\newblock \emph{arXiv preprint arXiv:2505.24066}, 2025.

\bibitem[Kim et~al.(2026)Kim, Ghosh, Pati, and Bandyopadhyay]{kim2026finite}
Jaehoan Kim, Indrajit Ghosh, Debdeep Pati, and Dipankar Bandyopadhyay.
\newblock Finite-rank logistic {G}aussian processes with exact likelihood for
  conditional density estimation.
\newblock \emph{arXiv preprint arXiv:2610.09452}, 2026.

\bibitem[Lenk(1991)]{lenk1991towards}
Peter~J. Lenk.
\newblock Towards a practicable {B}ayesian nonparametric density estimator.
\newblock \emph{Biometrika}, 78\penalty0 (3):\penalty0 531--543, 1991.

\bibitem[Leonard(1978)]{leonard1978density}
Tom Leonard.
\newblock Density estimation, stochastic processes and prior information.
\newblock \emph{Journal of the Royal Statistical Society: Series B
  (Methodological)}, 40\penalty0 (2):\penalty0 113--146, 1978.

\bibitem[Lindgren et~al.(2011)Lindgren, Rue, and
  Lindstr{\"o}m]{lindgren2011explicit}
Finn Lindgren, H{\aa}vard Rue, and Johan Lindstr{\"o}m.
\newblock An explicit link between {G}aussian fields and {G}aussian {M}arkov
  random fields: the stochastic partial differential equation approach.
\newblock \emph{Journal of the Royal Statistical Society: Series B (Statistical
  Methodology)}, 73\penalty0 (4):\penalty0 423--498, 2011.

\bibitem[Liu et~al.(2020)Liu, Ong, Shen, and Cai]{liu2020gaussian}
Haitao Liu, Yew-Soon Ong, Xiaobo Shen, and Jianfei Cai.
\newblock When {G}aussian process meets big data: A review of scalable {GP}s.
\newblock \emph{IEEE Transactions on Neural Networks and Learning Systems},
  31\penalty0 (11):\penalty0 4405--4423, 2020.

\bibitem[{Mansour L} et~al.(2022){Mansour L}, Seguin, Smith, and
  Zalesky]{seguin2022CSS}
Sina {Mansour L}, Caio Seguin, Robert~E. Smith, and Andrew Zalesky.
\newblock Connectome spatial smoothing ({CSS}): Concepts, methods, and
  evaluation.
\newblock \emph{NeuroImage}, 250:\penalty0 118930, 2022.

\bibitem[McCullagh and Nelder(1989)]{mccullagh1989generalized}
Peter McCullagh and John~A. Nelder.
\newblock \emph{Generalized Linear Models}.
\newblock Chapman and Hall, London, second edition, 1989.

\bibitem[M{\o}ller and Waagepetersen(2004)]{moller2004statistical}
Jesper M{\o}ller and Rasmus~Plenge Waagepetersen.
\newblock \emph{Statistical Inference and Simulation for Spatial Point
  Processes}.
\newblock Chapman and Hall/CRC, Boca Raton, 2004.

\bibitem[M{\o}ller et~al.(1998)M{\o}ller, Syversveen, and
  Waagepetersen]{moller1998log}
Jesper M{\o}ller, Anne~Randi Syversveen, and Rasmus~Plenge Waagepetersen.
\newblock Log {G}aussian {C}ox processes.
\newblock \emph{Scandinavian Journal of Statistics}, 25\penalty0 (3):\penalty0
  451--482, 1998.

\bibitem[Moyer et~al.(2017)Moyer, Gutman, Faskowitz, Jahanshad, and
  Thompson]{moyer2017continuous}
Daniel Moyer, Boris~A Gutman, Joshua Faskowitz, Neda Jahanshad, and Paul~M
  Thompson.
\newblock Continuous representations of brain connectivity using spatial point
  processes.
\newblock \emph{Medical Image Analysis}, 41:\penalty0 32--39, 2017.

\bibitem[Nocedal and Wright(2006)]{nocedal2006numerical}
Jorge Nocedal and Stephen~J. Wright.
\newblock \emph{Numerical Optimization}.
\newblock Springer, New York, second edition, 2006.

\bibitem[Papandreou and Yuille(2010)]{papandreou2010gaussian}
George Papandreou and Alan~L Yuille.
\newblock Gaussian sampling by local perturbations.
\newblock In \emph{Advances in Neural Information Processing Systems},
  volume~23, 2010.

\bibitem[Park and Friston(2013)]{park2013structural}
Hae-Jeong Park and Karl Friston.
\newblock Structural and functional brain networks: from connections to
  cognition.
\newblock \emph{Science}, 342\penalty0 (6158):\penalty0 1238411, 2013.

\bibitem[Rasmussen and Williams(2006)]{rasmussen2006gaussian}
Carl~Edward Rasmussen and Christopher K.~I. Williams.
\newblock \emph{Gaussian Processes for Machine Learning}.
\newblock MIT Press, Cambridge, MA, 2006.

\bibitem[Riihim{\"a}ki and Vehtari(2014)]{riihimaki2014laplace}
Jaakko Riihim{\"a}ki and Aki Vehtari.
\newblock Laplace approximation for logistic {G}aussian process density
  estimation and regression.
\newblock \emph{Bayesian Analysis}, 9\penalty0 (2):\penalty0 425--448, 2014.

\bibitem[Rue et~al.(2009)Rue, Martino, and Chopin]{rue2009approximate}
H{\aa}vard Rue, Sara Martino, and Nicolas Chopin.
\newblock Approximate {B}ayesian inference for latent {G}aussian models by
  using integrated nested {L}aplace approximations.
\newblock \emph{Journal of the Royal Statistical Society Series B: Statistical
  Methodology}, 71\penalty0 (2):\penalty0 319--392, 2009.

\bibitem[Saat{\c{c}}i(2012)]{saatci2012scalable}
Yunus Saat{\c{c}}i.
\newblock \emph{Scalable Inference for Structured {G}aussian Process Models}.
\newblock PhD thesis, University of Cambridge, 2012.

\bibitem[Simpson et~al.(2016)Simpson, Illian, Lindgren, S{\o}rbye, and
  Rue]{simpson2016going}
Daniel Simpson, Janine~B Illian, Finn Lindgren, Sigrunn~H S{\o}rbye, and
  H{\aa}vard Rue.
\newblock Going off grid: computationally efficient inference for
  log-{G}aussian {C}ox processes.
\newblock \emph{Biometrika}, 103\penalty0 (1):\penalty0 49--70, 2016.

\bibitem[St-Onge et~al.(2018)St-Onge, Daducci, Girard, and
  Descoteaux]{st2018surface}
Etienne St-Onge, Alessandro Daducci, Gabriel Girard, and Maxime Descoteaux.
\newblock Surface-enhanced tractography ({SET}).
\newblock \emph{NeuroImage}, 169:\penalty0 524--539, 2018.

\bibitem[Tokdar(2007)]{tokdar2007towards}
Surya~T. Tokdar.
\newblock Towards a faster implementation of density estimation with logistic
  {G}aussian process priors.
\newblock \emph{Journal of Computational and Graphical Statistics}, 16\penalty0
  (3):\penalty0 633--655, 2007.

\bibitem[Ubaru et~al.(2017)Ubaru, Chen, and Saad]{ubaru2017fast}
Shashanka Ubaru, Jie Chen, and Yousef Saad.
\newblock Fast estimation of $\mathrm{tr}(f(a))$ via stochastic {L}anczos
  quadrature.
\newblock \emph{SIAM Journal on Matrix Analysis and Applications}, 38\penalty0
  (4):\penalty0 1075--1099, 2017.

\bibitem[Wang et~al.(2025)Wang, Li, and Zhang]{wang2025riemannian}
Lu~Wang, Didong Li, and Zhengwu Zhang.
\newblock Riemannian diffusion kernel-smoothed continuous structural
  connectivity on cortical surface.
\newblock \emph{Imaging Neuroscience}, 3:\penalty0 IMAG.a.912, 2025.

\bibitem[Whittle(1963)]{whittle1963stochastic}
Peter Whittle.
\newblock Stochastic processes in several dimensions.
\newblock \emph{Bulletin of the International Statistical Institute},
  40\penalty0 (2):\penalty0 974--994, 1963.

\bibitem[Zalesky et~al.(2010)Zalesky, Fornito, Harding, Cocchi, Y{\"u}cel,
  Pantelis, and Bullmore]{zalesky2010whole}
Andrew Zalesky, Alex Fornito, Ian~H Harding, Luca Cocchi, Murat Y{\"u}cel,
  Christos Pantelis, and Edward~T Bullmore.
\newblock Whole-brain anatomical networks: does the choice of nodes matter?
\newblock \emph{NeuroImage}, 50\penalty0 (3):\penalty0 970--983, 2010.

\bibitem[Zhang et~al.(2022)Zhang, Daducci, He, Schiavi, Seguin, Smith, Yeh,
  Zhao, and O’Donnell]{zhang2022quantitative}
Fan Zhang, Alessandro Daducci, Yong He, Simona Schiavi, Caio Seguin, Robert~E
  Smith, Chun-Hung Yeh, Tengda Zhao, and Lauren~J O’Donnell.
\newblock Quantitative mapping of the brain’s structural connectivity using
  diffusion {MRI} tractography: A review.
\newblock \emph{NeuroImage}, 249:\penalty0 118870, 2022.

\end{thebibliography}

\appendix

\section{Spectral form of the SPDE precision}\label{sec: spectral precision}

Let $C_1$ be the lumped mass matrix and $G_1$ the stiffness matrix of the piecewise linear basis on one sphere, and let $C_1^{-1/2} G_1 C_1^{-1/2} = U \Lambda U^\top$ with $U$ orthogonal and $\Lambda = \diag(\lambda_1, \ldots, \lambda_{n_v})$, $0 = \lambda_1 \le \lambda_2 \le \cdots$. Since $C_1$ is diagonal, the product-domain matrices satisfy
\begin{align*}
C &= C_1 \otimes C_1 = (C_1^{1/2} \otimes C_1^{1/2})^2,
\\
G &= C_1 \otimes G_1 + G_1 \otimes C_1
\\
&= (C_1^{1/2} \otimes C_1^{1/2})\,(U \otimes U)\,(\Lambda \otimes I + I \otimes \Lambda)\,(U \otimes U)^\top (C_1^{1/2} \otimes C_1^{1/2}).
\end{align*}
Writing $V = C_1^{1/2} U \otimes C_1^{1/2} U$ and $\Sigma_\kappa = \kappa^2 I + \Lambda \otimes I + I \otimes \Lambda$, we obtain $K_\kappa = \kappa^2 C + G = V \Sigma_\kappa V^\top$ and $V^\top C^{-1} V = I$. Hence
\begin{equation*}
K_\kappa C^{-1} K_\kappa C^{-1} K_\kappa = V \Sigma_\kappa^3 V^\top,
\qquad
\big(K_\kappa C^{-1} K_\kappa C^{-1} K_\kappa\big)^{-1} = (\Phi \otimes \Phi)\,\Sigma_\kappa^{-3}\,(\Phi \otimes \Phi)^\top,
\end{equation*}
with $\Phi = C_1^{-1/2} U$. A draw $\mathrm{vec}(W) = (\Phi \otimes \Phi)\,\Sigma_\kappa^{-3/2}\,\mathrm{vec}(Z)$ with $\mathrm{vec}(Z) \sim \mathcal{N}(0, I)$ therefore has the SPDE precision up to scale, and in matrix form this is \eqref{eqn: spectral representation} up to the scale $c_\theta$ and the entry $(D_\theta)_{11}$.

Since $\Phi^\top C_1 \Phi = I$, the area-weighted average of the variance of $W$ over vertex pairs equals $\sum_{a,b} (\kappa^2 + \lambda_a + \lambda_b)^{-3} / A_S^2$, where $A_S = \mathrm{tr}(C_1)$ is the area of the sphere. The scale $c_\theta$ in \eqref{eqn: spectral scale} makes this average equal to $\tau^2$. The whole construction requires one symmetric eigendecomposition of size $n_v$ per mesh.

\section{Implementation details}\label{sec: implementation}

The linear systems of Fisher scoring and of the posterior draws are solved by conjugate gradients with a diagonal preconditioner that approximates the diagonal of $F$. Each Fisher scoring step is followed by a backtracking line search \citep[Chapter 3]{nocedal2006numerical}, which halves the step until $h_\theta$ decreases by at least $10^{-4}$ times the decrease predicted by its gradient. Fisher scoring stops when the decrease of $h_\theta$ falls below $10^{-6}$ relative or $10^{-3}$ absolute. It is initialized at $Z = 0$ and $\beta_b = \log\{n_b / \sum_{k,k'} a_k a_{k'}\}$, and warm started across successive values of $\theta$.

Conjugate gradients stop at relative residual $10^{-4}$ in both the Fisher scoring steps and the posterior draws. The stochastic Lanczos quadrature uses $4$ Rademacher probes and $30$ Lanczos steps, with the same probes for every $\theta$. For the real data, which are fitted with large $\tau$, we allow up to $60$ Fisher scoring steps and $400$ conjugate gradient iterations per step.

In the simulation, $(\log\kappa, \log\tau)$ is optimized by a pattern search with initial step $\log 1.35$, which is halved when no coordinate move improves the marginal likelihood, and at most $8$ evaluations. Results with a strict setting ($10^{-5}$ tolerances, $8$ probes, $40$ Lanczos steps, and Nelder--Mead with up to $40$ evaluations) agree to four decimals.

\section{Evaluation measures for the real data}\label{sec: real data measures}

Both measures score an estimated connectivity density on streamlines that were not used to fit it. For each scan $s$, that is, one subject at one time point, the streamlines are split at random into two halves. Under the model, by the thinning property of the Poisson process, the two halves are independent Poisson processes whose intensities are both half of the original one, so the normalized density estimated from one half targets the normalized density of the other half. Each method is fitted on one half, which gives the normalized mass
\begin{equation*}
\hat p_{s,r} = \frac{\int_{R_r} \hat L_s}{\sum_{r'} \int_{R_{r'}} \hat L_s}
\end{equation*}
of each region pair $R_r$, where the region pairs run over the triangle pairs of the three blocks and the normalization is joint over the blocks as in \eqref{eqn: intensity transformation to density}. For the proposed method, $\hat L_s$ is evaluated at the posterior mode.

The held-out log-score of scan $s$ is the average of $\log \hat p_{s, r(i)}$ over the streamlines $i$ of the other half, where $r(i)$ is the ico3 triangle pair that contains the endpoint pair of streamline $i$. Since this score estimates $\sum_r p_{s,r} \log \hat p_{s,r}$, where $p_{s,r}$ is the true mass, its difference between two methods estimates the difference of their Kullback--Leibler divergences from the true distribution of the region pairs. A difference of $\delta$ means that one method assigns, in the geometric mean over streamlines, a factor $e^{\delta}$ more probability to the region pair of a new streamline.

The cross-session log-score asks how well the estimate from one time point describes the scan at the other time point. Let $\hat p^{(t)}_{j,r}$ be the estimate of subject $j$ from the fitted half of the scan at time $t \in \{1, 2\}$, and let $N^{(t')}_{j,r}$ be the number of all streamlines of subject $j$ at the other time $t'$ in region pair $r$. Here the region pairs are those of the ico2 mesh, each the union of $16$ ico3 triangle pairs. The score is
\begin{equation*}
X^{t \to t'}_{j} = \frac{\sum_r N^{(t')}_{j,r} \log \hat p^{(t)}_{j,r}}{\sum_r N^{(t')}_{j,r}},
\end{equation*}
and we report the average of $X^{1 \to 2}_{j}$ and $X^{2 \to 1}_{j}$ over subjects. The score is high for an estimate that captures the connectivity reproduced at the other time point.

\end{document}